\documentclass[showkeys,showpacs,superscriptaddress]{revtex4}
\usepackage{amsmath}
\usepackage{amssymb}
\usepackage{graphicx}
\usepackage{color}

\begin{document}

\title{Thin-shell toroidal wormhole
}
\author{
Vladimir Dzhunushaliev
}
\email{v.dzhunushaliev@gmail.com}
\affiliation{
Department of Theoretical and Nuclear Physics, al-Farabi Kazakh National
University, Almaty 050040, Kazakhstan
}
\affiliation{
Institute of Experimental and Theoretical Physics, al-Farabi Kazakh National
University, Almaty 050040, Kazakhstan
}
\affiliation{
Institute of Physicotechnical Problems and Material Science of the NAS of the
Kyrgyz Republic, 265 a, Chui Street, Bishkek 720071, Kyrgyzstan
}

\author{Vladimir Folomeev}
\email{vfolomeev@mail.ru}
\affiliation{
Institute of Experimental and Theoretical Physics, al-Farabi Kazakh National
University, Almaty 050040, Kazakhstan
}
\affiliation{
Institute of Physicotechnical Problems and Material Science of the NAS of the
Kyrgyz Republic, 265 a, Chui Street, Bishkek 720071, Kyrgyzstan
}

\author{
Burkhard Kleihaus
}
\email{kleihaus@theorie.physik.uni-oldenburg.de}
\affiliation{
Institut f\"ur Physik, Universit\"at Oldenburg, Postfach 2503
D-26111 Oldenburg, Germany
}

\author{
Jutta Kunz
}
\email{kunz@theorie.physik.uni-oldenburg.de}
\affiliation{
Institut f\"ur Physik, Universit\"at Oldenburg, Postfach 2503
D-26111 Oldenburg, Germany
}


\begin{abstract}
We consider a topologically nontrivial thin-shell wormhole with a throat in the form of a  $T^2$ torus. It is shown that: (i)~such a wormhole is stable with respect to excitations of the throat;
(ii)~not all energy conditions are violated for such wormholes; (iii)~if any of the energy conditions is violated, this violation occurs only partially in some region near the throat, and in other regions the violation is absent. Also, we discuss the differences between spherical $S^2$ wormholes and toroidal $T^2$ wormholes under investigation.
\end{abstract}

\pacs{04.40.--b}

\keywords{torus, wormhole, thin-shell formalism}
\date{\today}

\maketitle

\section{Introduction}

The history of wormhole solutions begins with a Schwarzschild black hole for which (after the transformation to the Kruskal coordinates) there exists a nonstatic wormhole \cite{Einstein:1935tc,Kruskal:1959vx}.
The physical interest in wormholes is manifold, and it consists, in particular, in the fact that they (according to Wheeler's figurative terminology)
are ``mass without mass'' and ``charge without charge''
\cite{Wheeler:1955zz,Misner:1957mt}.
The existence of such objects would allow to have a
\emph{classical} model of a charge without central singularity possessing an asymptotically Coulomb field and finite total energy.
It is obvious that such a model would not be a model of an electron since the Pauli principle is not fulfilled for it and it is not clear how the electron's wave function
could be associated with such an object. Nevertheless, it should be noted that, at the moment, there is an assumption that wormholes can connect
entangled quantum objects (see, for example, Refs.~\cite{Maldacena:2013xja,Jensen:2013ora,Susskind:2017nto}).

The key point in obtaining wormhole solutions within Einstein's general relativity is using matter violating
the weak/null energy conditions (such matter is referred to as exotic matter).  Its presence allows the possibility of obtaining
compact configurations with a nontrivial spacetime topology -- traversable wormholes \cite{Thorne:1988}.
As exotic matter, one can take various types of matter. In particular, in the early pioneering works,
 massless ghost scalar fields  (i.e., fields with the opposite sign in front of the kinetic term of the scalar field Lagrangian density) have been used
 \cite{Ellis:1973yv,Bronnikov:1973fh} or scalar fields  with a Mexican hat potential were employed \cite{Kodama:1978dw,Kodama:1979}.
Traversable Lorentzian wormholes were further studied using matter fields
that would allow for such spacetimes
\cite{Kuhfittig:2010wz,lxli,ArmendarizPicon:2002km,Sushkov:2002ef,Lobo:2005us,Sushkov:2005kj}.
(For a general overview on Lorentzian wormholes, see the books of Visser and
Lobo \cite{Visser:1995cc,Lobo:2017oab}).

Consideration of wormholes has also been carried out within various modified theories of gravity. In that case
wormhole geometries can be theoretically constructed without involving exotic matter. In particular, this can be
$f(R)$ theories of gravity~\cite{Lobo:2009ip,DeBenedictis:2012qz,Harko:2013yb}, Einstein-Gauss-Bonnet gravity~\cite{Kanti:2011jz,Kanti:2011yv},
$f(R)$ theories of gravity coupled to the Born-Infeld electrodynamics~\cite{Bambi:2015zch}, $f(T)$ gravity~\cite{Jamil:2012ti}, in Eddington-inspired Born-Infeld gravity \cite{Shaikh:2015oha}, etc.

If one works within general relativity, there are the following distinctive features and problems typical for
\emph{most} wormhole solutions obtained so far:
(a) all such solutions are unstable~\cite{Shinkai:2002gv,Gonzalez:2008wd,Gonzalez:2008xk,Bronnikov:2011if,Bronnikov:2012ch,Torii:2013xba,Dzhunushaliev:2014bya}
(see, however, Ref.~\cite{Bronnikov:2013coa} where it is shown that for some exotic equations of state one can obtain stable wormholes);
(b) a static wormhole geometry requires the presence of exotic matter (at least near the throat).
In this connection, it would be of great interest to derive solutions possessing new properties,
possibly solving the present problems of wormhole solutions.

Most of the wormhole solutions obtained so far have a $S^2$ topological structure, where a minimal area surface (a throat)
has the topology of a 2-sphere. Another type of topology is exemplified by cylindrical wormholes (with the topology $S^1\times I$, where $S^1$ is a circle in the $(x,y)$ plane and $I$ is an interval along the $z$ axis) as
considered in \cite{Bronnikov:2009na} (for the asymptotically not flat case) and in \cite{Bronnikov:2018uje} (for the asymptotically flat case).
In the present paper we show that another type of wormholes with a
$T^2 = S^1 \times S^1$ topology is possible, where the throat has the topology of a 2-torus. We show that in this case important properties of the throat can change significantly. Thus we demonstrate that the simplest $T^2$ throat, obtained by cutting out the inner parts of a 2-torus belonging to two Minkowski spacetimes with a subsequent matching the obtained spaces along these two tori, is stable with respect to toroidal perturbations. We also show that for such wormholes not all energy conditions are violated, and for those which are violated, the violation occurs only in some region of the throat.

This permits us to speculate that the drawbacks of a throat associated with the need for violating the energy conditions might be related to its topology.
A $S^2$ throat is topologically trivial, while a $T^2$ throat is topologically nontrivial in the sense that not all curves can be contracted to a point.
This might possibly lead to changes in the stability properties of the throat.
However, a more complete study of the stability properties of such
$T^2$ wormholes is called for.

Our goal here is to study the possibility of existence of wormholes with a toroidal throat topology of the type $T^2 = S^1 \times S^1$. Wormholes of this kind have been considered by Gonzalez-Diaz in his pioneering paper~\cite{GonzalezDiaz:1996sr}, where he presents and analyzes a metric describing a tunnel between two spacetimes with topology other than spherical, and refers to such a wormhole as a ringhole. We will call such wormholes as toroidal or $T^2$ wormholes to emphasize the presence of a torus in the structure of the throat. We will then employ the thin shell formalism,
address the energy conditions and consider stability of these wormholes.

\section{$T^2 $ wormhole}

In the present paper we consider a toroidal throat
defined as a surface with a minimum cross section
having the topology of 2-torus $T^2$.
In doing so, we will study the simplest case when there are two copies
of flat Minkowski spacetime in which the inner parts of a torus are cut out and the resulting spaces are matched along these tori.
To describe such a spacetime, let us
introduce toroidal coordinates in a flat space:
\begin{eqnarray}
	x &=& a \frac{\sinh \alpha \cos \varphi}{\cosh \alpha - \cos \beta} ,
\label{tor_coord_x}\\
	y &=& a \frac{\sinh \alpha \sin \varphi}{\cosh \alpha - \cos \beta} ,
\label{tor_coord_y}\\
	z &=& a \frac{\sin \beta}{\cosh \alpha - \cos \beta} ,
\label{tor_coord_z}
\end{eqnarray}
where $0 \leq \alpha < \infty$, $- \pi \leq \beta \leq \pi$,
$0 \leq \varphi \leq 2 \pi$. The torus with $\alpha = \infty$ is a degenerate
torus, i.e., a circle. The center of a torus with $\alpha = 0$ is located at
infinity.

In toroidal coordinates, the flat spacetime metric is
\begin{equation}
	ds^2 = dt^2 - \left(  \frac{a}{\cosh \alpha - \cos \beta} \right)^2
	\left(
		d \alpha^2 + d \beta^2 + \sinh^2 \alpha d \varphi^2
	\right).
\label{tor_metric}
\end{equation}
Here $a$ is some parameter. The section $t, \alpha = \text{const.}$  describes the $T^2$ torus and $\beta, \varphi$ are the coordinates on the torus.

The coordinate surfaces are:
\begin{itemize}
\item the torus $T^2$ by $\alpha = \text{const.}$,
	\begin{equation}
		\left(
			\sqrt{x^2 + y^2} - a \coth \alpha
		\right)^2 + z^2 = \left(
			\frac{a}{\sinh \alpha}
		\right)^2 ;
	\label{coord-torus}
	\end{equation}
\item	the $S^2$ sphere by $\beta =\text{const.}$,
	\begin{equation}
		\left(
			z - a \cot \beta
		\right)^2 + x^2 + y^2 = \left(
			\frac{a}{\sin \alpha}
		\right)^2 ;
	\label{coord-sphere}
	\end{equation}
\item
	half-planes by $\varphi =\text{const.}$,
	\begin{equation}
		\frac{x}{y} = \cot \varphi .
	\label{coord-varphi}
	\end{equation}
\end{itemize}
The toroidal wormhole can be obtained as follows: We cut out a torus that is located
near a circle (a degenerated torus with $\alpha = \infty$) and join it with
the analogous construction from another spacetime.

\section{Thin-shell wormhole}

\subsection{Thin-shell formalism}

In this section we closely follow the book~\cite{Visser:1995cc}.
We seek a solution to the Einstein equations with a delta-like energy-momentum tensor of a thin shell of matter
\begin{eqnarray}
	R_{\mu \nu} - \frac{1}{2} g_{\mu \nu} R &=& 8 \pi G T_{\mu \nu} ,
\label{3-a-10}\\
	T_{\mu \nu} &=& \delta(\alpha) S_{\mu \nu},
\label{3-a-20}
\end{eqnarray}
where $\alpha$ is the Gaussian coordinate normal to the throat and
 $S_{\mu \nu}$ denotes the surface stress-energy. The tensor $S_{\mu \nu}$ is defined as follows:
 \begin{equation}
	S_{\mu \nu} = - \frac{1}{8 \pi G} \left(
		\kappa_{\mu \nu} - \kappa h_{\mu \nu}
	\right) .
\label{3-a-30}
\end{equation}
Here $\kappa_{\mu \nu}$ is the discontinuity in the second fundamental form
$K_{\mu \nu}$,
\begin{equation}
	\kappa_{\mu \nu} = K^+_{\mu \nu} - K^-_{\mu \nu},
\label{3-a-40}
\end{equation}
and $h_{\mu \nu}$ is the projection tensor defined as follows:
\begin{equation}
	h_{\mu \nu} = g_{\mu \nu} - n_\mu n_\nu,
\label{3-a-50}
\end{equation}
where $g_{\mu \nu}$ is the metric tensor and $n_\mu$ is the unit vector normal
to the shell. 
The upper indices
$()^{+, -}$ refer to quantities at
the two sides of the shell, 
respectively.

The second fundamental form is defined as follows:
\begin{equation}
	K_{\mu \nu}^{\pm} = \frac{1}{2} \left(
		\nabla_\mu^{\pm} n_\nu + \nabla_\nu^{\pm} n_\mu
	\right),
\label{3-60}
\end{equation}
where $\nabla_\mu^{\pm}$ is the covariant derivative.

\subsection{Thin-shell toroidal $T^2$ wormhole from flat space}

Consider the simplest $T^2$ wormhole obtained in the following manner. According to the form of the toroidal metric~\eqref{tor_metric}, the surface
 $\alpha = \alpha_0 = \mathrm{const.}$ is a torus [see Eq.~\eqref{coord-torus}].
 So we take two flat spaces and cut in both cases the inner part of a torus at  $\alpha = \alpha_0$.
 After that, we match these spaces along these tori.

Now we have the following metric:
\begin{equation}
	g^\pm_{\mu \nu} = \mathrm{diag} \left\{
		-1, \left(
			\frac{a}{\cosh ( \tilde \alpha \pm \alpha_0) - \cos \beta}
		\right)^2,
		\left(
			\frac{a}{\cosh ( \tilde \alpha \pm \alpha_0) - \cos \beta}
		\right)^2,
		\left(
			a \frac{ \sinh ( \tilde \alpha \pm \alpha_0)}{\cosh ( \tilde \alpha \pm
			\alpha_0) - \cos \beta}
		\right)^2 	
	\right\} .
\label{3-b-20}
\end{equation}
Here $\tilde \alpha + \alpha_0$ and $\tilde \alpha \geq 0$ correspond to one
half of the wormhole and $\tilde \alpha - \alpha_0$ and $\tilde \alpha \leq 0$
-- to the other half of the wormhole.

For this metric, there are the following unit vectors normal to the center of the throat (where $\tilde \alpha = 0$):
\begin{equation}
	n^\pm_\mu = \left(
		0, \frac{a}{\cosh ( \tilde \alpha \pm \alpha_0) - \cos \beta} , 0 , 0
	\right) .
\label{3-b-30}
\end{equation}
Using the metric \eqref{3-b-20} and the vector  \eqref{3-b-30},
one has the following second fundamental form
\begin{equation}
	K^\pm_{\mu \nu} = a \begin{pmatrix}
		0 & 0	&	0	&	0 \\
		0 & 0	&	\frac{\sin \beta}{2 \left( \cosh \alpha_0 - \cos \beta \right)^2 }
		&	0 \\
		0 &\frac{\sin \beta}{2 \left( \cosh \alpha_0 - \cos \beta \right)^2 }	
		&	\mp \frac{\sinh \alpha_0}{\left( \cosh \alpha_0 - \cos \beta \right)^2}	
		&	0 \\
		0 & 0	&	0	&	\mp \sinh \alpha_0
		\frac{-1 + \cos \beta \cosh \alpha_0}{\left( \cosh \alpha_0 - \cos \beta
		\right)^2}
	\end{pmatrix}
\label{3-b-40}
\end{equation}
and the discontinuity tensor
\begin{equation}
	\kappa_{\mu \nu} = \mathrm{diag} \left\{
		0, 0,
		- \frac{2 a \sinh \alpha_0}{\left( \cosh \alpha_0 - \cos \beta \right)^2},
		- 2 a \sinh \alpha_0
		\frac{-1 + \cos \beta \cosh \alpha_0}
		{\left( \cosh \alpha_0 - \cos \beta \right)^2}
	\right\} .
\label{3-b-50}
\end{equation}
Calculating the  surface stress-energy  according to the formula
\eqref{3-a-30}, we have
\begin{equation}
	S_{\bar \mu \bar \nu} = \frac{1}{4 \pi G a} \mathrm{diag} \left\{
		\frac{-1 + \cosh \alpha_0 \cos \beta +
		\sinh^2 \alpha_0}{\sinh \alpha_0} , 0,
		- \frac{-1 + \cos \beta \cosh \alpha_0}{\sinh \alpha_0} ,
		- \sinh \alpha_0
	\right\} .
\label{3-b-60}
\end{equation}
[Note that henceforth we work with tetrad components of the  surface stress-energy (the corresponding tensor indices have bars).] On the other hand, the surface stress-energy can be represented as
\begin{equation}
	S_{\bar \mu \bar \nu} = \mathrm{diag} \left\{
		\sigma , 0,
			- \theta_\beta , - \theta_\varphi
	\right\},
\label{3-b-70}
\end{equation}
where $\sigma$ is the surface energy density and $\theta_{\beta, \varphi}$ are the principal surface tensions. Thus we have
\begin{eqnarray}
	\sigma &=& \frac{1}{4 \pi G a}
	\frac{-1 + \cosh \alpha_0 \cos \beta + \sinh^2 \alpha_0}{\sinh \alpha_0},
\label{3-b-80}\\
	\theta_\beta &=& \frac{1}{4 \pi G a}
	\frac{-1 + \cos \beta \cosh \alpha_0}{\sinh \alpha_0},
\label{3-b-90}\\
	\theta_\varphi &=& \frac{1}{4 \pi G a}
	\sinh \alpha_0  .
\label{3-b-100}
\end{eqnarray}

Let us calculate the scalar curvature of the toroidal $T^2$ throat, whose metric is
\begin{equation}
	dl^2 = \left( \frac{a}{\cosh \alpha - \cos \beta} \right)^2
		\left(
			d \beta^2 + \sinh^2 \alpha d \varphi^2
		\right) .
\label{tor_metric_throat}
\end{equation}
Using this metric yields
\begin{equation}
	R_{\text{throat}} = \frac{2}{a^2} \left(
		- 1 + \cosh \alpha \cos \beta
	\right)~.
\label{scal_curv}
\end{equation}

\section{Energy conditions}

In this section we address the energy conditions for the throat described above.

The null energy condition asserts that for any null vector $k_\mu$
\begin{equation}
	T_{\mu \nu} k^\mu k^\nu \geq 0 ,
\label{4-10}
\end{equation}
or in terms of the principal pressures $p_i$
\begin{equation}
	\epsilon + p_i \geq 0.
\label{4-20}
\end{equation}
In our case we have the following inequalities rewritten in terms of the principal surface tensions (minus the principal pressures):
\begin{eqnarray}
	\sigma - \theta_{\beta} &\geq& 0 ,
\label{4-30}\\
	\sigma - \theta_{\varphi} &\geq& 0 .
\label{4-40}
\end{eqnarray}

\begin{figure}[t]
\centering
\includegraphics[width=0.5\linewidth]{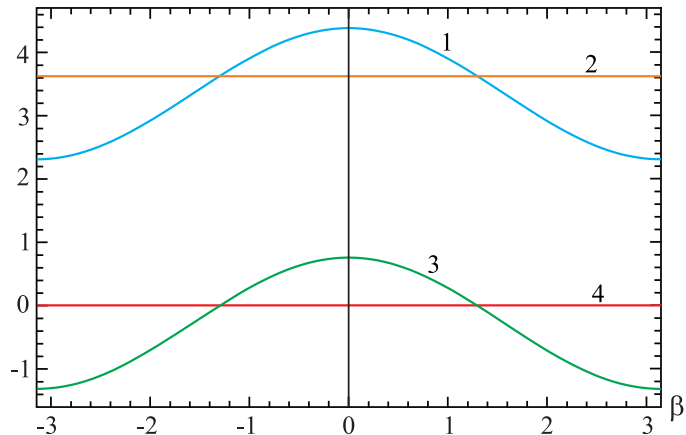}
\caption{The profiles of the energy conditions for $\alpha_0=2$. The curve 1 corresponds to
$\sigma(\beta)$,
	the curve 2 -- to $(\tilde \sigma(\beta) - \tilde \theta_{\beta}(\beta))$,
	the curve 3 -- to $(\tilde \sigma(\beta) - \tilde \theta_{\varphi}(\beta))$,
	the curve 4 -- to $(\tilde \sigma(\beta) - \tilde \theta_{\beta}(\beta) -
	\tilde \theta_{\varphi}(\beta))$.}
\label{enegyconditions}
\end{figure}

The weak energy conditions asserts that for any timelike vector $V_\mu$
\begin{equation}
	T_{\mu \nu} V^\mu V^\nu \geq 0 .
\label{4-50}
\end{equation}
In terms of the principal surface tensions $\theta_i$ this gives
\begin{equation}
	\sigma \geq 0 \quad \text{ and } \quad \sigma - \theta_i \geq 0.
\label{4-60}
\end{equation}
The strong energy condition asserts that for any timelike vector $V_\mu$
\begin{equation}
	\left(
		T_{\mu \nu} - \frac{1}{2} g_{\mu \nu} T
	\right) V^\mu V^\nu \geq 0 .
\label{4-70}
\end{equation}
In terms of the surface tensions we then have
\begin{equation}
	\sigma - \theta_i \geq 0 \quad \text{ and } \quad
	\sigma - \sum\limits_{i} \theta_i \geq 0.
\label{4-80}
\end{equation}

It is seen from the above that in order to analyse possible violations of the energy conditions for the system under consideration
one needs to keep track of the behavior of the following expressions:
\begin{equation} \sigma(\beta),\quad
\sigma(\beta) -  \theta_{\beta, \varphi}(\beta),\quad
 \sigma(\beta) -  \theta_\beta(\beta) -
\theta_\varphi(\beta)
\label{4-90}.
\end{equation}
Taking into account Eqs.~\eqref{3-b-80}-\eqref{3-b-100} and introducing the
dimensionless variables $	\tilde \sigma = 4 \pi G a \sigma,
	\tilde \theta_{\beta, \varphi} = 4 \pi G a \theta_{\beta, \varphi}$,
one obtains
\begin{equation}
	\tilde \sigma(\beta) - \tilde \theta_{\beta}(\beta)\equiv\sinh \alpha_0,
	\quad
	\tilde \sigma(\beta) - \tilde \theta_{\varphi}(\beta) \equiv
	\frac{-1+\cos\beta \cosh\alpha_0}{\sinh\alpha_0}, \quad
	\tilde \sigma(\beta) - \tilde \theta_\beta(\beta)
	- \tilde{\theta}_\varphi(\beta)\equiv 0.
\label{4-91}
\end{equation}
Taking into account that $\alpha_0$ is assumed to be positive, it is seen from
the second expression that its right-hand side will inevitably become negative
for some values of $\beta$. This corresponds to the fact that the violation of
the null/weak energy condition occurs (at least on some part of the
torus). Moreover, one can see from \eqref{3-b-80} that for small positive
$\alpha_0$ the surface energy density $\sigma(\beta)$ will also become negative
for some $\beta$'s. This corresponds to the violation of  the weak energy
condition.

Examples of the corresponding graphs are shown in Fig.~\ref{enegyconditions}.
From this figure, one can see that the energy condition
$(\sigma - \theta_\varphi)$ is violated in the regions of the torus with
$-\pi \leq \beta < - \arccos (1/\cosh\alpha_0)$ and $(\arccos 1/\cosh\alpha_0) < \beta \leq \pi$, but is not violated in the region with
$-\arccos (1/\cosh\alpha_0) \leq \beta \leq \arccos (1/\cosh\alpha_0)$.

\section{Stability}
\label{stability}

A stability analysis of the  $T^2$ throat is performed exactly as it is done for a $S^2$ throat of Ref.~\cite{Visser:1995cc}.
To study the stability of the system, let us consider a throat whose linear sizes can change in time:
\begin{equation}
	g^\pm_{\mu \nu} = \mathrm{diag} \left\{
		-1, \left(
			\frac{a}{\cosh ( \tilde \alpha \pm \alpha_0(t)) - \cos \beta}
		\right)^2,
		\left(
			\frac{a}{\cosh ( \tilde \alpha \pm \alpha_0(t)) - \cos \beta}
		\right)^2,
		\left(
			a \frac{ \sinh ( \tilde \alpha \pm \alpha_0(t))}
			{\cosh ( \tilde \alpha \pm \alpha_0(t)) - \cos \beta}
		\right)^2 	
	\right\} .
\label{5-10}
\end{equation}
The equations of motion of the center of the throat can be found from the  covariant conservation of the energy-momentum tensor, $\nabla_\mu T^{\mu \nu} = 0$.
It gives two equations
\begin{eqnarray}
	S^{\mu \nu} n_\nu &=& 0,
\label{5-20}\\
	\frac{1}{2} S^{\mu \nu} \left(
		K^+_{\mu \nu} + K^-_{\mu \nu}
	\right)  &=& 0 .
\label{5-30}
\end{eqnarray}
From calculations of the surface stress-energy  $S_{\mu \nu}$ and of the second fundamental
forms $K^\pm_{\mu \nu}$ with the metric \eqref{5-10}, we have the following equations:
 \begin{eqnarray}
	\frac{\sinh \alpha_0 \sqrt{
		\left(
			\cos \beta - \cosh^2 \alpha_0
		\right)^2 + a {\dot \alpha_0}^2
	}}{\left(
				\cos \beta - \cosh^2 \alpha_0
			\right)^2} \dot \alpha_0 &=& 0 ,
\label{5-40}\\
	- \frac{\sinh \alpha_0}{\cos \beta - \cosh^2 \alpha_0} {\dot \alpha_0}^2
	&=& 0 ,
\label{5-50}\\
	- 4 \frac{-1 + \cosh \alpha_0 \cos \beta + \sinh \alpha_0}
	{\sinh \alpha_0 \left(
				\cos \beta - \cosh^2 \alpha_0
			\right)} \ddot \alpha_0	&=& 0.
\label{5-60}
\end{eqnarray}
They have a trivial solution $\alpha_0 = \mathrm{const.}$ This means that the
throat under consideration is stable against toroidal perturbations. In
Eq.~\eqref{5-60}, we took into account Eqs.~\eqref{5-40} and \eqref{5-50}.

Here a subtle point should be noted. In the metric \eqref{tor_metric}, there is the parameter  $a$ and, at first glance, to study the stability, one has to consider the case of  $a(t)$.
But in this case the metric \eqref{tor_metric} changes accordingly in time. This means that one has to consider more general perturbation of the throat, that should be done in future studies of a  $T^2$ wormhole.

\section{Discussion and conclusions}

We have obtained the simplest model of a wormhole with a topologically nontrivial toroidal throat. This throat is obtained using the
thin-shell formalism by matching two copies of flat Minkowski spacetimes along 2-tori. Such a throat differs in principle from a throat possessing a $S^2$ topology. The study of the properties of the  $T^2$ wormhole indicates that they evidently depend considerably on the topology of the throat.
For a more comprehensive study of the differences of the properties of  $T^2$ and $S^2$ wormholes
it seems necessary  to construct  more realistic  $T^2$  wormhole solutions.
For example, one could go beyond the thin-shell formalism used here.
But this poses, unfortunately, a considerably more difficult technical problem. In particular, the equations describing a $T^2$ wormhole with the metric of the type
\eqref{tor_metric} will then be partial differential equations,
that certainly complicates the procedure of obtaining their solutions.

In our opinion, the most interesting questions in future studies of $T^2$ wormholes are: (i) their stability; (ii) the necessity of using exotic matter to construct a $T^2$ wormhole;
(iii) the possibility of the existence of wormholes filled with spinor fields, which, in our opinion, is necessary for considering the problem of the
connection of entangled quantum electrons by a wormhole, according to the ideas of Refs.~\cite{Maldacena:2013xja,Jensen:2013ora,Susskind:2017nto}.

Here, working within the thin-shell formalism, we have shown that
\begin{itemize}
  \item There exist wormholes with a throat having the topology of 2-torus $T^2$ and connecting two flat Minkowski spacetimes.
  \item At least in some regions of such a throat, there should occur the
  violation of the null/weak energy conditions. Also, the sign of the surface
  energy density depends on $\alpha_0$: for small $\alpha_0$ the surface energy
  density is negative ($\sigma < 0$) and for large $\alpha_0$ it is positive ($\sigma > 0$) (recall that small (large)
  $\alpha_0$ correspond to large (small) tori $T^2$).
  \item The throat appears to be (absolutely) stable against toroidal perturbations.
\end{itemize}

\section*{Acknowledgments}
V.D. and V.F. gratefully acknowledge support provided by Grant No.~BR05236322
in Fundamental Research in Natural Sciences by the Ministry of Education and Science of the Republic of Kazakhstan. We are grateful to the Research Group
Linkage Programme of the Alexander von Humboldt Foundation for the support of this research and also would like to thank the Carl von Ossietzky University of
Oldenburg for hospitality while this work was carried out.
B.K. and J.K. would like to acknowledge support from the DFG Research Training Group 1620 {\sl Models of Gravity}
and support by the COST Action CA16104.

\end{document}